Chapter  #

# KM AND WEB 2.0 METHODS FOR PROJECT-BASED LEARNING.

# MESHAT: A MONITORING AND EXPERIENCE SHARING TOOL

CHRISTINE MICHEL [1], ÉLISE LAVOUÉ [2]
[1] *Université de Lyon, INSA Lyon, Laboratoire LIESP, F-69621 Villeurbanne, France*
[2] *Université Jean Moulin Lyon 3, IAE Lyon, Équipe de recherche MAGELLAN, Groupe SICOMOR, Lyon, France*

Abstract:   Our work aims to study tools offered to students and tutors involved in face-to-face or blended project-based learning activities. To better understand the needs and expectations of each actor, we are especially interested in the specific case of project management training. The results of a course observation show that the lack of monitoring and expertise transfer tools involves important dysfunctions in the course organisation and therefore dissatisfaction for tutors and students (in particular about the acquisition of knowledge and expertise). To solve this problem, we propose a personalised platform (according to the actor: project group, student or tutor), which gives information to monitor activities and supports the acquisition and transfer of expertise. This platform is based on Knowledge Management (KM) and Web 2.0 concepts to support the dynamic building of knowledge. KM is used to define the learning process (based on the experiential learning theory) and the way the individual knowledge building is monitored (based on metacognitive concepts). Web 2.0 is used to define the way the experience is shared. We make the hypothesis that this approach improves the acquisition of complex skills (e.g. management, communication and collaboration), which requires a behavioural evolution. We aim to make the students become able 'to learn to learn' and evolve according to contexts. We facilitate their ability to have a critical analysis of their actions according to the situations they encounter.

## 1.     INTRODUCTION

Project-based learning is often applied in the case of complex learning (i.e. which aims to make students acquire various linked skills or develop their behaviour). In comparison to traditional learning, this type of learning relies on co-development, collective responsibility and cooperation. Students are the principal actors of their learning. A significant enrichment arises from their activity, both for them and all the other students. A consequence of this approach is the segmentation of the class into sub-grouped projects, monitored by tutors. We generally observe that the coordination and harmonisation of tutors' activities are extremely difficult to operate when each group works autonomously, on different subjects and in real and varied environments (for example enterprises). It is even more difficult when the project is conducted over a long period (more than four weeks). In this context, the perception of individuals' and groups' activity is also very difficult, especially if no technical support for information and communication is used. Finally, the implementation of project-based learning in engineering schools, universities or professional training do not benefit from all its capacities (Thomas & Mengel, 2008). Indeed, this learning should implement an educational model based on the Kolb's cycle (Cortez et al., 2009), composed of four phases: concrete experience, reflective observation, abstract conceptualisation and active experimentation. However, it is often action (via the articulation conceptualisation-experimentation) which is favoured to the detriment of concrete experience and reflective observation (Thomas & Mengel, 2008).

To better understand the type of tool necessary to improve this training, we have studied a project management training course (Michel & Prévot, 2009). This course is supported by a rich and complex organisation, especially for tutors that we detail in Section 2. We have used KM methods to identify all the problems encountered by students and tutors and identify the following three main problems.
1. Difficulties in students acquiring some skills (e.g. project management organisation, use of monitoring tools and groupwork) and autonomy.
2. A lack of information so that tutors can monitor and evaluate students individually and by group.
3. A lack of tutors' communication and coordination so that they develop their expertise, knowledge and competences.

In Section 3, we study existing tools which can help to solve these problems, especially monitoring and experience sharing tools. We then observe that no existing tool could solve all these problems on its own. Therefore we propose a new tool named MEShaT (Monitoring and Experience Sharing Tool) before finally concluding with the future directions offered by this work.

## 2.     CASE STUDY: A PROJECT MANAGEMENT TRAINING COURSE

### 2.1     The course organisation

The course is composed of a theoretical presentation of the principles and methods of project management and their practical application to a project (called 'PCo' for 'Collective Project') carried out by groups (12 groups of 8 students which answer to different industrial needs). Envisaged by Patrick Prévôt (Michel & Prévot, 2009), the project management course lasts six months and corresponds to an investment of approximately 3000 students' working hours per project. The instructional objectives are to acquire hard competences (e.g. knowing how to plan the project (Gantt's chart), project management, managing resources, controlling quality) and soft competences (e.g. social competences of collaboration and communication, empathy, consideration of others, leadership). The pedagogical team (see Figure 1) is composed of 24 tutors (a technical and a management tutor per group), two managers (technical and management) in charge of the coordination of the technical and management tutors' activities, one teacher who presents the theoretical concepts and one director responsible for the organisation of the training of all groups.

The project is composed of four phases:
1. November: answer to the call for tender (formalisation of the client's requirements).
2. December: elaboration of a master plan (means, tools and organisation of the team project), definition of tools to drive the project (dashboard) and rules to test the deliverables quality (rules of receipt).
3. January to March: development of a product or a study.
4. Until mid-April: delivery of a technical report which describes the product and management report (a project closure report which is an analysis, from the student's point of view, of the flow and problems of the project). The project is closed by one dramatised presentation in front of all the actors of the project.

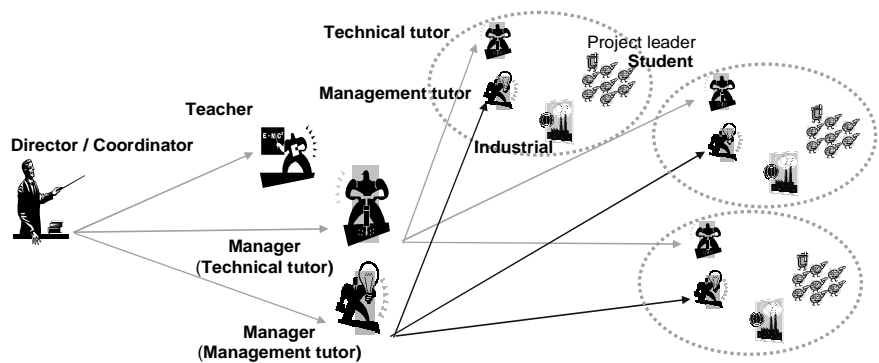

Figure 1 - Pedagogical team and course organisation

The course has been designed according to the experiential learning theory, well-known in KM, and is based on the expanded learning circle proposed by Berggren and Söderlund (2008). This circle is based on the learning circle developed by Kolb & Kolb (2005), which consists of concrete experience, reflective observation, abstract conceptualisation and active experimentation, in combination with the different learning styles. Berggren and Söderlund (2008) expanded this model and propose a social twist of experiential learning (see Figure 2):
- The processes of *articulation* and *reflection* allow for the abstraction of knowledge.
- The processes of *investigation* and *enaction* contribute to the social character of knowledge and the diffusion of experience developed within the educational programmes.

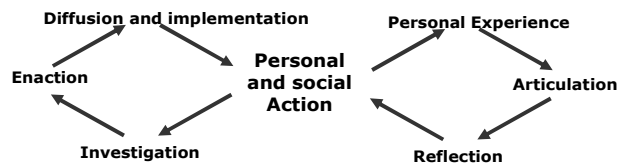

Figure 2 – The expanded learning circle (Berggren & Söderlund 2008)

In the framework of the project, the *personal experience* of the student is a result of the education process constructed by following the right circle of the model (see Figure 2). The *articulation* phase corresponds to debriefing discussions and debate driven by the tutors (one face-to-face discussion per week). The teacher presents the 'soft' and 'hard' concepts to students during the course. During the realisation of the project, students discuss 'soft' concepts with their management tutor and 'hard' concepts with the technical tutor so as to analyse and understand them. Tutors therefore play the role of animators. This work is strongly linked with the *reflection* phase (especially *reflective observation*s). Reflective observation can occur in a tacit way after these discussions with tutors or in a more formal way by realising the management report or other deliverables. Students choose, alone or according to tutors' instructions, the *personal and social actions* useful for the project or relevant according to the teaching objectives. It helps them build a unique experience, not well formalised by the teaching team. In this case, reflection is articulated with the *personal and social actions* and helps to apply 'hard' and 'soft' concepts and to build competences. It is also an occasion for each individual to express their personal experience.

Another characteristic of our project design is to promote, on the one hand, the processes of *investigation* and *enaction*, and, on the other hand, the process of *diffusion*. The courses in project management usually consist of realising a well-known project (a case study). In our case, the *investigation* process is emphasised by the fact that students have to solve an industrial problem without a predefined solution. It provides a real challenge that facilitates the construction of knowledge and improves *enaction*. The *diffusion* process is realised in the form of dramatised representations. Students present their good/bad practices and their feelings and judgments about the training and the tutors. These representations take part in a KM diffusion process, between the project team, the teaching team and the department. They also aim to support the reflection and conceptualisation processes necessary for students to realise the experience they gain by working in a group.

This combination of activities go with an evolution of behaviour in terms of skills (management, communication, collaboration and all 'soft' competences) and natural reactions (to be able to learn how to learn and to evolve in surprising or unknown situations) by supporting the students' capacity for self-critical analysis. This capacity mainly results from the training activities carried out with the tutors. Indeed the tutors play various roles which depend on the type of skills the students have to acquire. According to Garrot's taxonomy (Garrot et al., 2009), for the acquisition of soft skills, tutors are social catalysts (by creating a friendly environment to incite students to participate), intellectual catalysts (by asking questions and inciting students to discuss and to criticise), 'individualisers' (by helping every student to overcome their difficulties, to estimate their needs, difficulties, preferences) and 'autonomisers' (by helping students to regulate their learning and to acquire autonomy). For the acquisition of hard skills, tutors are relational coaches (by helping students to learn how to work in a group and to become a leader), educationalists (by redirecting groups' activities in a productive way, clarifying points of methodology, supplying resources), content experts (by answering questions on the course contents), evaluators (by evaluating students and groups' productions and participation) and 'qualimetrors' (by measuring and giving feedback on the quality of the course).

Tutors monitor a unique and non-reproducible project. They work with students most of the time face-to-face and no organisation, communication or capitalisation tool is proposed. For example, no specific tool is currently proposed to the tutors for the monitoring of students' activities or for their evaluation. The appreciation of students' activity is made in an implicit way, according to the number and the quality of face-to-face student–tutor interactions. In terms of communication and coordination, each tutor works individually with their group and does not communicate much with the other tutor of the same group (management or technical) in order to have a complete vision of the group's activity.

## 2.2    The observed problems

The observation methodology is adapted from the Method for Knowledge System Management (MASK) approach (Benmahamed et al., 2005). This method, starting from documents produced by an organisation and talks with actors, allows the modelling of complex industrial systems by identifying and inter-relating various concepts: product, actor, activity, rules and constraint. Each concept is defined on a card; the Information, Constraint, Activity, Rule, Entity (ICARE) cards describe any *object* precisely intervening in the process. The Reuse, Improve and Share Experiment (RISE) cards describe any *problem* occurring during the process and specify the contexts, suggested solutions or recommendations. The elements described in the ICARE and RISE cards are organised overall in a *chart,* which shows their interrelationships (the method is completely described in (Michel & Prévot, 2009)).

For this research, we analyse results from RISE cards. The observation data are various experience feedbacks from students and tutors and were collected by 62 students in the fifth year of engineering school. The observed students are 23 males and 18 females who are between 22 to 25 years old. Thirty-eight of them have carried out the project management course the previous year, three of them are currently 'project leaders'. Observation consists of direct feedbacks made by interview of the course director, of six tutors and of three students currently 'project leaders' and by self-observation for the other 38 students. Indirect feedbacks are based on various groups' experience and analyses expressed in their 'management report', which is one of the projects deliverables. Twenty-four management reports have been considered (each one relating to the experience of a group). By this observation we have identified 36 different types of problems described in RISE cards.

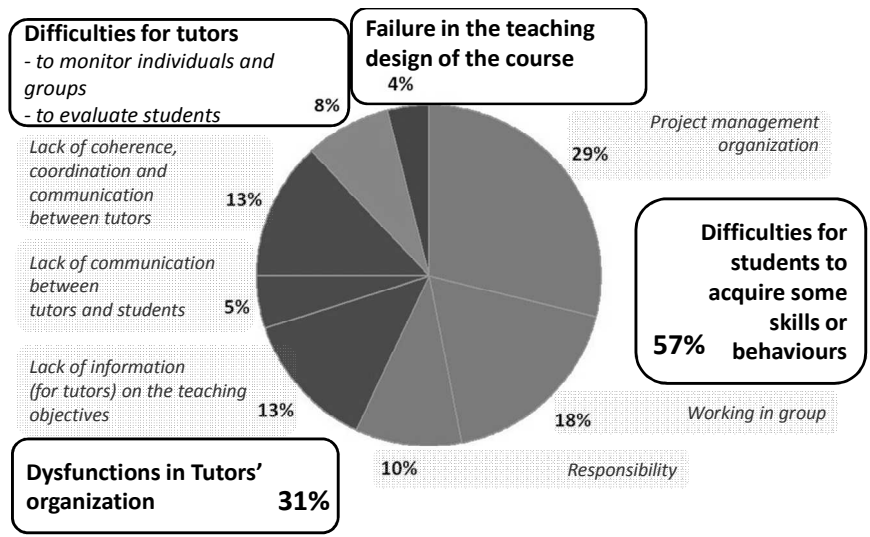

Figure 3 – Observed problems

The type and frequency of the observed problems are presented in Figure 3. The majority of cards (57%) relate to a problem with the management of the teamwork by the team itself. More precisely, 29% relate to a lack of project management skills, 18% relate to difficulties working in group, 10% relate to problems with some students who think they are not responsible enough. Meanwhile, 31% of the problems concern tutors' activity and impact on the teaching organisation of the project. Indeed, 13% concern a lack of coherence, coordination and communication between tutors, which involves problems of information diffusion. For example, the instructions given to the project groups were described as ambiguous or contradictory. About 5% concern a lack of communication between tutors and students or a lack of presence of some tutors; 13% concern a lack of information for tutors on the teaching objectives or on the knowledge and skills they have to teach to students. Indeed, students feel alone when they have to learn using some tools or when they have to apply theoretical project management concepts. Students sometime do not understand the role tutors play and the help they can bring them. Moreover, 8% of the problems concern failure in the teaching design of the course (not enough time to work, a not adapted calendar and too short timing for the deliverables). Finally, many groups and tutors express the same problem concerning the monitoring of individuals' or groups' activity and students evaluation (4% of the problems). The students express a feeling of injustice concerning the individual evaluation because the notation is the same for all members of a project (with about + or -2 points according to their investment), even if the students are involved more or less than the others. All the tutors also express their difficulties in evaluating the students individually. These difficulties are explained by the intuitive and tacit character of the evaluations, by the lack of traceability of students' actions, and by the lack of discussion with their colleagues.

It is possible to partially solve problems concerning the course design and the course organisation by changing the timing and the teachers' and coordinators' responsibilities. Nevertheless many problems remain and most of them are directly or indirectly bound to tutors' activity. That is why we aim to help tutors, on the one hand, to monitor and to evaluate the students and the groups and, on the other hand, to exchange information, coordinate and develop their skills and expertise. Although the pedagogical context is not distance learning, we hope to benefit from using tools to support this activity. In the next part, we study knowledge management and Web 2.0 tools which are suitable for our case. We focus on monitoring tools and expertise sharing tools.

## 3. TOOLS TO SUPPORT LEARNING ACTIVITIES

In this part, we detail existing tools to help tutors to monitor students' activities and to communicate with the other tutors. We study how these tools can help tutors and solve the problems identified in the previous part. We finally show that none of them answer all the needs and therefore we develop our own tool.

### 3.1 Monitoring tools

Many tools have been developed to support tutors in the monitoring of distant and synchronous students' individual activities. ESSAIM (Després, 2003) gives a global view of a student's progress in the course and tutors have a perception of the activity with reference to the path, the actions and the productions of each student. FORMID (Guéraud & Cagnat, 2006) offers a tutor interface with a global view of a class during a session (e.g. students' login, their progress in the course) or a zoom-in on a precise course stage (successfully validated or not by the class, by a student or by a group of students so as to identify their difficulties). These tools work in a synchronous environment with automatically generated tracks. They are thus only meant for tutors and do not offer the possibility for students to regulate their learning for a long period. Furthermore, they are not meant for asynchronous learning situations for which tutors need information on students' activities over a long period.

Other tools are meant to help tutors to monitor asynchronous activities and entice students towards their autonomy or to regulate their learning by determining themselves the state of their progress in the course. Croisières (Gueye, 2005) offers services which individually support students in their learning progress and assist them in autonomy situation. Students select their learning activities according to their objectives and learning strategies. Reflet (Després & Coffinet, 2004) is a tool meant for showing the state of progress of a student or a class. It supplies information to the tutors who monitor the students in distance training and to the students who have feedback on their progress with regard to the learning objectives and the other students. Students determine their state of progress in the course with regard to the tasks they have to carry out and tutors can deny students the validation of some of their tasks.

There are also tools to monitor the activities of groups, not simply individuals. SIGFAD (Mbala et al., 2005) offers a support for actors' interactions in restricted groups (8–15 persons) in distance learning. It helps tutors to hold the groups, to boost them and indeed to conduct the course well. The interaction statistics allow one to model and to show the collaboration into groups, to estimate the group's life and evolution. SIGFAD supplies three main categories of estimations: at the level of the group (present, absent or still persons, the state of the group with regard to the realisation of the activities), at the level of individuals (their productivity in terms of the realisation of activities and their sociability which indicates their level of communication with the other members of the group) and at the level of the activity (level of realisation of an activity by all participants). TACSI (Laperrousaz et al., 2005) offers more specifically a perception of the individual students' activity into the activity of their group. It distinguishes the perception of students' activity in an individual task (individual productions), the perception of students' activity in a collective task (their contributions in the collective activities and their contributions to the discussions) and the perception of students' situation in the group dynamics (social behaviour and sociometric status). The LCC (Learning to Collaborate by Collaborating) collaborative activity software (Cortez et. al, 2009) is used for teaching and measuring teamwork skills using technologically supported face-to-face collaborative activities. LCC allows seven variables to be measured: the first variables measure the activity score (i.e. the group's efficiency in performing the task assigned), while the last variables measure teamwork (corresponding to core components (skills) of teamwork like team orientation (TO), team leadership (TL), monitoring (MO), feedback (FE), back-up (BA) and coordination (CO)). Communication has not been included in the measurable variables.

    The individual and collective indicators for the monitoring of students and project groups offered by these tools are relatively well adapted to our context. We especially adopt those proposed within the LCC framework (Cortez et al., 2009) for the development of our own monitoring tool. However, the course which interests us does not use instrumented activity and thus does not allow using automatically collected tracks of students' activity which is why we have to think about other ways of collecting information on their activities.

    The tools which help students to acquire autonomy incite them to evaluate their progress in the course, according to the tasks they have achieved and those they have to achieve. However, these tools are not adapted because they do not help students to build an individual reflection neither on the relevance of the knowledge they acquire and the modalities of this acquisition nor on their behavioural changes. These self-regulatory processes are individual and mainly result from the activities carried out with the tutors. We think it useful (Michel & Prévôt, 2009) to support these processes by using a metacognitive tool (Azevedo, 2007) which takes into account students' point of view of cognition (e.g. activating prior knowledge, planning, creating sub-goals, learning strategies), metacognition (e.g. feeling of knowing, judgment of learning, content evaluation), motivation (e.g. self-efficacy, task value, interest, effort) and behaviour (e.g. engaging in help-seeking behaviour, modifying learning conditions, handling task difficulties and demands).

    All the tools studied in this part are exclusively centred on students' activity and help neither students nor tutors to have reflections on their activity. In our context, in which the roles played by tutors are extremely varied, it is essential to have a base structuring this reflection. For example, Berggren & Söderlund (2008) propose to use a 'learning contract' defined as 'a number of fairly simple questions, such as: What do I want to learn? How will I learn this? Who can give support? When can I start? How will I know that I have learned? How will others realise that I have learned?' This contract could be useful not only for students but also for tutors.

    Furthermore, all the tools do not help tutors to understand or interpret what they observe. They supply useful information for tutors but this information is rather quantitative than qualitative and thus does not allow the evaluation of the quality of the contributions or productions, or to explain students' behaviour neither individually nor inside the group. These tools can be useful for tutors only if they know how to use them, how to interpret the supplied information and how to react effectively and in an adapted way. Finally, these tools address every tutor individually and do not allow them to coordinate at the level of monitoring of the same

project group and to exchange on their activity so as to acquire more expertise. All of which is why we go on to study in the next section the tools which support exchanges between tutors to allow them to help each other and to develop their skills.

## 3.2 Experience sharing tools

The results of a previous study (Michel et al., 2007) about tools supplied to tutors shows that they do not have adapted tools to exchange or formalise their experience as allowed, for example, by Knowledge Based Systems (KBS) or experience booklets (Kamsu Foguem et al., 2008). Furthermore, we observed that tutors are rather structured in a hierarchical way within the organisation and do not have coordination tools or dedicated spaces for meeting between peers.

To compensate for a lack of training and formal help, Communities of Practice (CoPs) of tutors emerge. Web technologies (e.g. forums, blogs, wikis) have allowed the emergence of online CoPs (Cuthell, 2008; Pashnyak & Dennen, 2007). CoPs gather tutors together in an informal way because of the fact that they have common practices, interests and purposes (i.e. to share ideas and experiences, build common tools, and develop relations between peers). Members exchange information, help each other to develop their skills and expertise and solve problems in an innovative way. They develop a community identity around shared knowledge, common approaches and established practices and create a shared directory of common resources (Wenger, 1998; Garrot-Lavoué, 2009). The use of technology does allow the accumulation of exchange, but these are relatively unstructured and not contextualised. Web tools such as blogs, mailing lists, chat and email, allow discussions without building concrete knowledge (only forums bring a slightly higher degree of explicit emergence, thanks to the spatial representation as discussion threads which highlights relations between messages).

Numerous works aim to answer the question by supplying tutors with tools to support specific activities. Some tools work through member participation and sociability, for example by offering a virtual 'home' like the Tapped In environment (Schlager & Fusco, 2004), others by supporting collaboration between members like CoPe_it! (Karacapilidis & Tzagarakis, 2007). Other tools favour the creation of contextualised resources and contextual search facilities such as the learning environment doceNet (Brito Mírian et al,. 2006). However, all these environments either favour sociability (engaging members to participate) to the detriment of the reification of the produced resources, or they favour the accumulation and indexation of contextualised resources, but to the detriment of sociability and member participation.

We have developed the TE-Cap platform (Garrot-Lavoué, 2009) so as to support a good structuralisation of the information without decreasing member participation (for example communication). Indeed, the tutors have discussions by way of contextualised forums: they associate tags with the discussions to describe the context. These tags are subjects of a tutoring taxonomy, shown in an interactive and evolutionary way (the tutors can propose new subjects for the taxonomy). This platform, associated with a monitoring tool, could answer our needs of knowledge and skills acquisition and capitalisation about the realisation of tutors' activity and about the use of the monitoring tools.

## 4. A PLATFORM FOR TUTORS AND STUDENTS

We have designed a customised platform called MEShaT (see Figure 4). It proposes different interfaces according to the learning actor: a project group, a student or a tutor. Every interface consists of the following.
1. A monitoring tool (on the form of a dashboard) which helps the concerned actor to have a global view of their activity.
2. A publication tool which allows the spread of their experience.

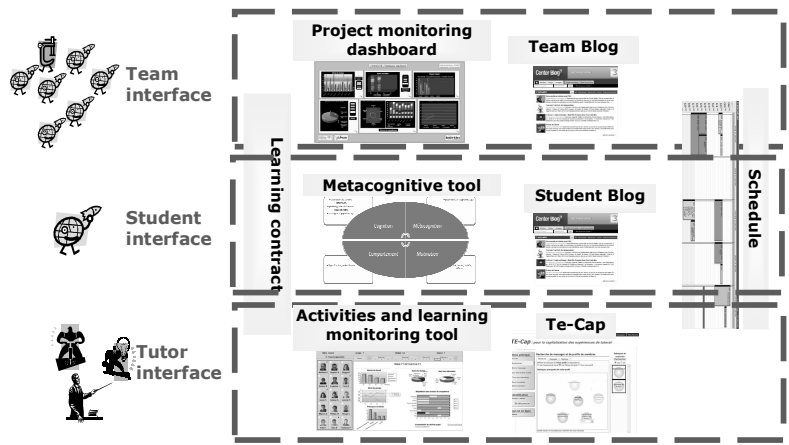

Figure 4 - MEShaT: Monitoring and experience sharing tool for project-based learning

Three dashboards are offered; two for students (one to monitor the progress of their project and the other one to monitor their own learning process) and one for tutors (to monitor students' and groups' activities and students' learning).

- The project monitoring dashboard is a project management tool meant for the group and shows various indicators: the group's frame of mind (e.g. motivation, satisfaction, relationship with the client), the Gantt diagram, tasks to realise and the percentage of realisation, the working time of each member, the deliverables to produce and the delays. This tool is dedicated to the group leader for the project steering, to the members to situate themselves regarding the others and to express themselves. The indicators provide information to students for the metacognitive processes described below and to tutors for the monitoring of teamwork.
- The metacognitive tool takes into account students' individual point of view of their cognition, metacognition, motivation and behaviour so as to build reflexive indicators. Concerning cognition, students evaluate themselves in relation to the target competencies in project management (hard and soft) as well as to the ones necessary for the project realisation. They define the planning, the sub-goals and the learning strategies required to acquire these competencies. Regarding metacognition, students express their feeling about competencies and knowledge acquisition (level, form, context, judgment). Students more precisely

    describe their motivation about their self-efficacy, the value and the interest of the tasks and the required effort. They also formalise their behaviour by explaining how they engage in help-seeking strategy and the way they handle task difficulties and demands. We consider that it is important to reflect all these indicators to students so as to help them to build an individual reflection on the relevance of the knowledge they acquire, on the modalities of this acquisition and on their behavioural changes.
- The activities and learning monitoring tool is meant for tutors and shows information on the individual students' activity and the groups' activity thanks to indicators such as the group orientation, leadership, monitoring, feedback and coordination. These indicators are built thanks to the information given by students in the individual and group dashboards described above. Tutors therefore have access to all the information on student and group activities and can intervene when needed. The history helps them to understand the individual and group processes, to intervene with the students in an adapted way and to assess the students' work.

    The publication tools are blogs and TE-Cap.
- Blogs (one per student and one per group) are spaces where students can freely describe, for example, the realisation contexts of their actions and their frame of mind. These blogs help the group members and the tutors to understand the project context, to explain the value of some indicators (as delays or the group's frame of mind) and so to anticipate or to solve problems more quickly.
- TE-Cap is offered to tutors to allow the emergence of a CoP composed of all the tutors who monitor a project. The indexation model is built on three main subjects, corresponding to the different types of expertise required for tutors: (1) their roles and tasks; (2) the project calendar (so as to coordinate); and (3) the specific progress of every group. By exchanging, tutors will acquire expertise on their roles and knowledge on their application ground. TE-Cap can be considered as an expertise transfer tool.

    A fixed section shows information accessible by all the actors: the schedule and the learning contract. The schedule helps students and tutors, of the same or different groups, to coordinate their activities. The learning contract defines simple questions for students such as: 'What do I want to learn? How will I learn this? Who can give support? When can I start? How will I know that I have learned? How will others realise that I have learned?' These questions are defined at the start of the project and are used to focus students' attention on the educational objectives throughout the project. This contract could be useful not only for students but also for tutors. Tutors can also refer to these kinds of questions to refocus on their roles. It is a means to coordinate tutors who have to implement the same educational means.

    The information on the dashboards can be modified by their owner(s) and are not visible for everybody. Students can modify their blog and their individual dashboard by means of a data entry interface. The groups' dashboard is updated by the project leader, using individual information. Leaders confirm the data and decide what is published on the blog. The tutors' dashboard is directly updated by them and automatically updated according to the information entry on the groups' and students' interfaces. Tutors also contribute directly to the CoP. Tutors have access to the groups' and students' interfaces. The project leaders have no access to the individual dashboards of their group members. The learning contract cannot be modified during the course progress. It is updated at the end of the project, according to the events which were related on blogs and on TE-Cap.

MEShaT is meant for the complex educational context of project-based learning, using the Kolb's learning process. The metacognitive tool, the blogs, TE-Cap and the learning contract, favour the reflexive observation and concrete experience phases of the Kolb's cycle, the monitoring tools help action phases (conceptualisation and experimentation). Moreover, MEShaT solves some of the problems identified in Section 2.2. Monitoring tools and blogs facilitate groupwork, group cohesion and the professionalism of students by making the consequences of their acts more tangible and by informing them. Metacognitive tools and blogs help students to acquire knowledge and reinforce their motivation (by a better understanding of what they have to do and why they do it). If these phenomena do not naturally appear, tools will help the tutors to make them emerge. Indeed, MEShaT reinforces the tutor-student link by allowing the continuous monitoring of the knowledge acquisition process. It also helps tutors to assume some of their roles, like their roles of relational coach and social catalyst (concerning groupwork or leadership), their role of intellectual catalyst (by asking precise and conceptualised questions to incite students to discuss or ask critical questions) and their roles of expert and pedagogue. Moreover, the association of Te-Cap with the learning contract offers tutors a space for refining or developing their expertise.

## 5. CONCLUSION AND FUTURE DIRECTIONS

Our work aims to study how KM methods and Web 2.0 tools can be useful in face-to-face or blended project-based learning activities. We propose to make use of them to design a monitoring and expertise transfer tool proposed to tutors and students. To understand better the needs and expectations of each actor, we are especially interested in the case of project management training. Indeed, this type of learning is complex since it has for an objective the acquisitions of soft and hard knowledge and relies on rich and varied social organisations. In the first part of this article we described a course that has been designed according to the experiential learning theory and based on an expanded learning circle. We then expose the observed problems, like the lack of monitoring and expertise transfer tools, which involve important dysfunctions in the course organisation and therefore dissatisfaction for tutors and students (in particular about the acquisition of knowledge and expertise). The study of existing tools highlights two points:

1. There is no tool which helps both tutors and students.
2. There are no clear strategies proposed to acquire, transfer and capitalise on the actors' experience.

Indeed, studied tools do not offer metacognitive functions, formal or informal publication tools (such as knowledge books or blogs) or tools to support CoP.

Therefore, to solve this problem, we propose to associate personalised monitoring tools (one for the project group, one for the student and one for the tutor) with tools for the transfer of experience and the acquisition of knowledge. Regarding the monitoring: the 'team feedback' is a dashboard for the project management, the 'student feedback' is a metacognitive tool and the 'tutor feedback' is a monitoring tool for individuals' and groups' activity. The tool for the acquisition of knowledge considers two types of knowledge: the acquired experience is formalised in a kind of knowledge book called a 'learning contract', the experience being acquired is revealed and capitalised in blogs (for students and project groups) and within a CoP supported by TE-Cap (for tutors). We describe their articulation in a platform: MEShaT. This platform is dedicated to project management education but can also be used to support different types of face-to-face project-based learning activities. Indeed, all the phases of the Kolb's cycle are taken into account. Furthermore, it supports the acquisition of various experiences: those of the individuals (students and tutors) and those of the social organisations (project group, CoP of tutors). Our future work will consist of testing this platform over a long time so as to experimentally validate our hypotheses. We will also observe how the actors (students, tutors and course designer) appropriate this type of technology and how they participate in the redefinition of their roles.

## 6.  ACKNOWLEDGEMENT

The author would like to thank René Peltier, director of Airbus KM service (Toulouse) until 2007, for his help in the adaptation of KM methods for our context.